# Imaging Techniques for Relativistic Beams: Issues and Limitations

Alex H. Lumpkin and Manfred Wendt

Fermilab[*] – Instrumentation Department
Batavia, IL 60510 – USA

Characterizations of transverse profiles for low-power beams in the accelerators of the proposed linear colliders (ILC and CLIC) using imaging techniques are being evaluated. Assessments of the issues and limitations for imaging relativistic beams with intercepting scintillator or optical transition radiation screens are presented based on low-energy tests at the Fermilab A0 photoinjector and are planned for the Advanced Superconducting Test Accelerator at Fermilab.

## 1  Introduction

One of the basic parameters to be characterized for an accelerated electron beam is the transverse beam size (and corresponding emittance). In the case of the International Linear Collider (ILC) the range of beam energies would ultimately go up to 5, 15, and even 250 GeV with beam sizes from 300 μm down to <10 μm, respectively [1]. There are smaller sizes in the vertical plane. However, much can be learned by using relativistic beams at sub-GeV energies such as at the Fermilab A0 Photoinjector (A0PI) in regard to fundamental issues and limitations of the conversion mechanisms and optical systems. We present here aspects of scintillators and optical transition radiation (OTR) screens that are used as intercepting techniques for the tune-up or low-intensity beam operation and should be applicable over a wide range of energies. As will be shown there are several issues on screen resolution or OTR polarization and point spread functions (PSFs) that must be properly addressed in order to determine successfully the actual beam size and profile. In addition, there is a possibility of beam instabilities such as the longitudinal-space-charge-induced-microbunching (LSCIM) instability that currently plagues the OTR diagnostics from 150 MeV to 14 GeV in the LCLS accelerator at SLAC [2]. This effect has also been observed in compressed bright beams in linacs at APS/ANL, FLASH, and Elettra [3-5]. The basic ILC-like pulse train is 3.2 nC per micropulse at 3 MHz in a 1-ms macropulse which is repeated at up to 5 Hz. Each micropulse is to be compressed to about 300 μm, or 1-ps-sigma bunch length at 40 MeV in the injector area before entering the cryomodules which contain eight 9-cell cavities. The Advanced Superconducting Test Accelerator (ASTA) currently under construction at Fermilab will generate such beams at near-GeV scale by using 3-4 cryomodules [6]. The tune-up beam will be comprised of up to 100 micropulses, and the spacing may be adjusted somewhat flexibly by selecting pulses in the drive laser for the photoinjector. This beam will be characterized by the imaging techniques with the intercepting screens, and there will be tests to see how many micropulses and with what charge one can robustly operate. Non-intercepting techniques will be applied to the high-power beam.

---

## 2 Beam-size Imaging Considerations of Accelerated Beam

A basic particle beam imaging system includes:
 -a **conversion mechanism**: (scintillator, optical or x- ray synchrotron radiation (OSR or XSR), optical transition radiation (OTR), Cherenkov radiation (CR), undulator radiation (UR), and optical diffraction radiation (ODR),
-**optical transport** (windows, lenses, mirrors, filters, polarizers),
-**imaging sensor** such as a CCD, CID, CMOS camera with or without image intensifier and/or cooling,
-**video digitizer** (built in or external), and an
-**image processing software.**

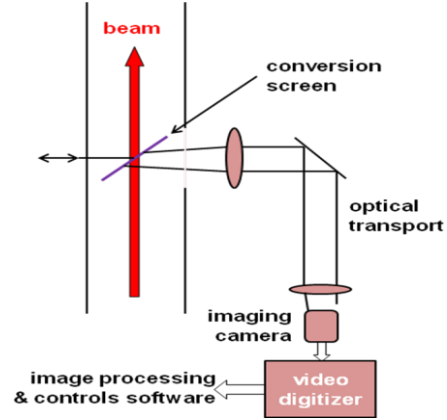

Figure 1: Schematic of beam-imaging system.

We then have to identify corrections to consider in our analysis of the beam image. The system related ones are: YAG:Ce powder and crystal screen resolution, OTR polarization effects, OTR point spread function, camera calibration factor, and finite slit size (if applicable). The accelerator and beam-related effects include the beta star term in the dispersive plane of a spectrometer and the macropulse blurring effects due to RF power or phase slew on beam size, energy spread, and beam divergence in OTR images that sum over many micropulses.

Uncorrelated terms are treated as a quadrature sum (see Lyons' textbook [7]) which contribute to the observed image size (Obs) including the actual image size (Act), YAG screen effects (YAG), camera resolution (Cam), and finite slit width (Slit) as shown in Eq. 1. In addition there can be macropulse effects and OTR polarization effects.

$$Obs^2 = Act^2 + YAG^2 + Cam^2 + Slit^2 \qquad (1)$$

 and solving for the actual beam size , we have

$$Act = \sqrt{Obs^2 - YAG^2 - Cam^2 - Slit^2} \qquad (2)$$

A series of experiments has been performed at the A0PI facility which is shown schematically in Fig. 2. The imaging cross stations are indicated as X# and most of the work was done at X3, X5, X23, X24, and the prototype station indicated. The facility operates with a photo-cathode RF gun followed by a superconducting L-band 9-cell cavity generating final beam energies of 13-15 MeV, with micropulse charges of 250 to 1000 pC [8].

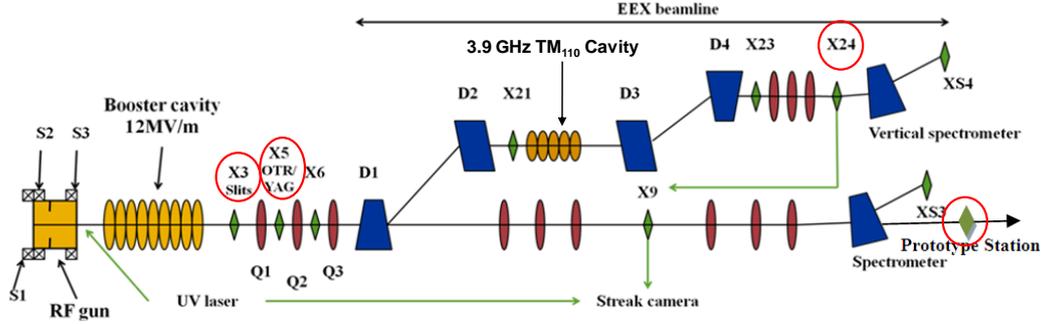

Fig. 2: A schematic of the A0PI facility with PC rf gun, superconducting booster cavity, diagnostics cross stations, the spectrometers, and EEX beamline.

2.1 Beam profiling with YAG:Ce Scintillator Screens

YAG:Ce powder screens used at the A0 Photoinjector had nominally a 5-µm grain size and were coated at 50-µm thickness on various metal 1-mm-thick substrates of Al or SS. In the A0PI arrangement the scintillator material was on the front surface of the substrate, and oriented at $45^0$ to the beam direction. The powder screens were kindly provided by K. Floettmann (DESY). Observed characteristics include the response time of about 80 ns FWHM, and there have been reports of saturation of the mechanism for incident electron beam areal charge densities ~10 fC/µm$^2$. This latter effect can cause a charge dependence of the observed image size in addition to the low-charge, screen-resolution limit. The initial comparison tests of the powder screens and OTR were done at X5. As shown in Table 1 the scintillator-based sizes are insensitive to the linear polarizer while the OTR x size is reduced by 23 µm out of 125 with the vertical polarizer. The deduced powder resolution term for this case is 80±20 µm using the polarized OTR as the reference size, and the average of three separate measurements is 60±20 µm.

| Screen type | No. of bunches | X5 linear polarization | Fit σ (pixels) | X size (µm) |
|---|---|---|---|---|
| OTR | 10 | none | 5.49±0.05 | 124.5 |
|  | 10 | vertical | 4.47±0.09 | 101.0 |
| YAG:Ce | 1 | none | 5.67±0.05 | 128.7 |
|  | 1 | vertical | 5.71±0.04 | 129.6 |

Table 1: Comparison of OTR and YAG:Ce screens at X5.

These powder screens were replaced by 100-µm thick single crystal YAG:Ce screens oriented normal to the beam followed by a 45 degree mirror. A summary of various tests of powder samples and single-crystal YAG:Ce is shown in Fig. 3. It is obvious that the resolution term for powder screens is thickness dependent and much larger than the grain size. It is also clear that the 100-μm thick single crystal normal to the beam provides better resolution than a powder screen of similar thickness. The material and screen orientation are given in the label near each datum [9].

## 2.2 OTR Imaging

The fundamental OTR mechanism occurs when a charged particle beam transits the interface between two media. The approaching charge and the induced image charge in the second medium may be treated as a collapsing dipole with the consequent emission of radiation, i.e. OTR. The yield is limited to about 1 visible photon per 1000 electrons incident, but they are emitted in the few-fs time scale as opposed to the slower 80-ns scintillation process in the previous section. The radiation is emitted around the angle of specular reflection for backward radiation and around the angle of the beam direction in the forward direction for high gamma beams. For an oblique incidence such as 45 degrees, backward OTR is emitted at 90 degrees to the beam direction as shown at the upper right of Fig. 4. This geometry is compatible with most accelerator beam profiling stations.

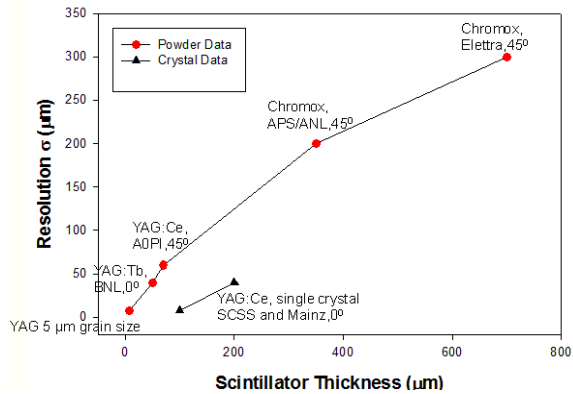

Fig. 3: A comparison of deduced resolution terms for powder screens and YAG:Ce crystals.

### 2.2.1 OTR Basics

The angular distribution pattern is annular with an opening angle of $1/\gamma$, where $\gamma$ is the Lorentz factor, as shown in Fig. 5. The peak intensity goes roughly as $1/\gamma^2$ and the spectral function as $1/\lambda^2$. The visibility of the central minimum depends on the beam divergence and is therefore related to beam transverse emittance. This visibility feature for OTR from a single foil is usable for divergence sensitivity down to about 10% of $1/\gamma$.

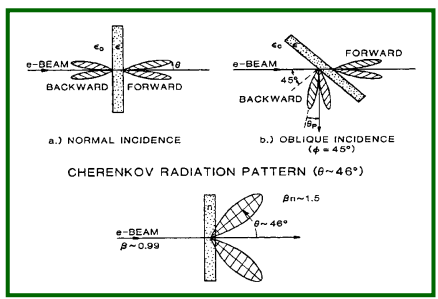

Fig. 4: A schematic of the OTR emission at the boundary of two media.

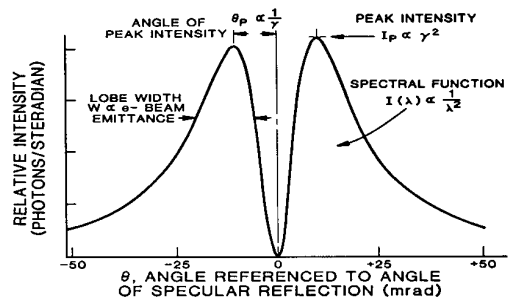

Fig. 5: A schematic of the basic OTR angular-distribution pattern and the dependence on beam parameters.

### 2.2.2 OTR Polarization and PSF Effects

During the course of our experiments with linear polarizers placed in the optical transport to the camera at the prototype station, we observed the OTR beam image size was smaller when we used the perpendicular polarization component relative to the beam dimension as shown in Fig. 6. The total OTR image is at the upper left, and the vertically polarized image is at the upper right. The fits to the projected x profiles gave sigma values of 66.8 ± 0.3 µm and 55.1 ± 1.1 µm, respectively. This effect at the 15-20% level at 55 µm we felt should not be ignored and further investigations are planned.

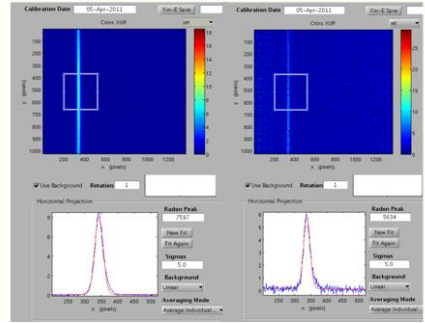

Fig. 6: Comparison of the OTR image (left) with the perpendicularly polarized OTR component narrower image (right). The projected profiles are below the images.

One possible explanation was to consider the OTR point-spread function that had been identified in the past by Castellano and Verzilov [10]. Basically, one convolves the OTR single electron angular distribution function with the $J_1$ Bessel function for diffraction from a point source as given in Eq. 3. The function argument involves $\theta_{max}$, $\gamma$, and $\varsigma = k\, R_i /M$ (where $k$ is the wave number, $R_i$ is the lens radius, and $M$ is the optical magnification). In this case one actually obtains an annular PSF at the few-micron level using visible light.

$$f^2(\theta_m, \gamma, \varsigma) = [\int_0^{\theta_m} \frac{\theta^2}{\theta^2 + \gamma^{-2}} J_1(\varsigma\theta) d\theta]^2 \qquad (3)$$

In their calculations they assumed a lens aperture of 100 mrad and calculated the total OTR PSF to be about 12 λ FWHM. They do calculate different projected profiles for the two polarization components which when convolved with the actual beam size would, in principle, give slightly different observed beam sizes. The effect due to the beam energy is negligible.

As an illustration of this, two cases for $E$=14.3 MeV, $M$=1, $\theta_{max}$=0.010 rad, λ=500 nm, and initial sigmas of 10 and 50 µm are shown in Fig. 7. The convolutions of total OTR and horizontally polarized OTR with horizontal and vertical projections with the Gaussian profiles are shown. For these input conditions we see ~10% effects at 50 µm, and 120 % effects at 10 µm. In the experiment we have about a 12-μm image size reduction at 55 μm using the perpendicular component compared to the 3-μm-reduction modeled result.

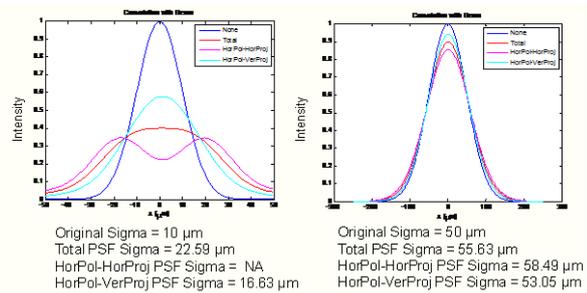

Fig. 7: A comparison of the OTR PSFs convolved with two Gaussian beams with sigma = 10 µm (left) and 50 µm (right).

### 2.2.3 Microbunching Instability and COTR

One of the recent `developments in diagnostics for compressed bright beams is the identification of the LSCIM instability and the appearance of dominating coherent OTR (COTR) signals [2,3]. Since this effect is attributed to noise fluctuations in the beam as it transports through the accelerator, the observed effects are random in spatial distribution and their local intensities preclude simple beam profile measurements. The effect is described by Ratner et al. [11], and the broad band nature of the gain is shown in Fig. 8 for the nominal LCLS case of a 3-keV slice energy spread. We have superimposed the CCD camera response curve and the incoherent OTR spectral distribution on the plot to illustrate the relationships.

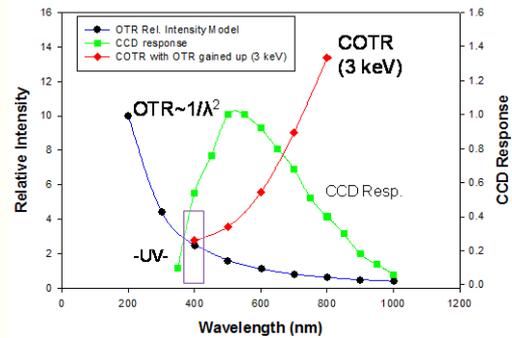

Fig. 8: A comparison of the spectral dependence of incoherent OTR and LSCIM COTR with the CCD response.

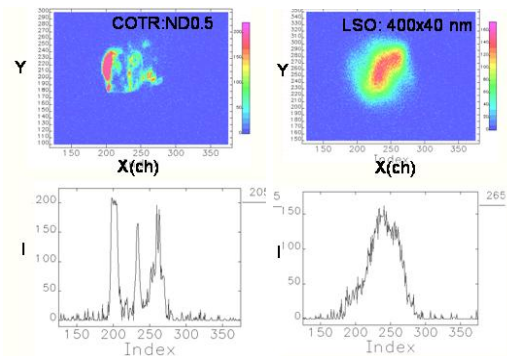

Fig. 9: A comparison of the COTR image (left) with the filtered LSO:Ce image (right). The projected x profiles are below each image, respectively.

It has been demonstrated as shown in Fig. 9 that by choosing the violet spectral region (such as indicated by the rectangle centered at 400 nm in Fig. 8), one can reduce the LSCIM COTR and still have some OTR signal. This can be made even more advantageous by using a scintillator that emits in the violet regime such as LSO:Ce at 415 nm. In addition, options to image in the ultraviolet down to 200 nm or even in the EUV appear feasible.

## 3  Future tests at ASTA

With the commissioning of the ASTA facility our techniques will be evaluated ultimately with beam energies up to the GeV scale. Technical progress includes the first cryomodule installation in the tunnel as shown in Fig. 10. In regard to the imaging stations, the prototype developed with RadiaBeam Technologies is shown in Fig. 11. We will use the normal incidence geometry and YAG:Ce crystals before the chicane and probably the LYSO:Ce crystals that emit near 415 nm after the chicane. The geometry will minimize the depth-of-focus aspects compared to the former A0PI 45-degree geometry, and the single crystal will avoid the larger powder resolution term. We also will have the OTR screen option, and we will use two linear polarizers oriented orthogonally and selectable in a filter wheel to provide the preferred polarization component. We will evaluate the OTR PSF effects and adjust the optics accordingly. It is expected that we will use the OTR screen at 45 degrees to the beam, and we will adjust the optical focus from the scintillator plane to this latter z position. We plan to mitigate any moderate microbunching instability COTR effects by using the 400 x 40 nm bandpass filter with the LYSO:Ce crystal.

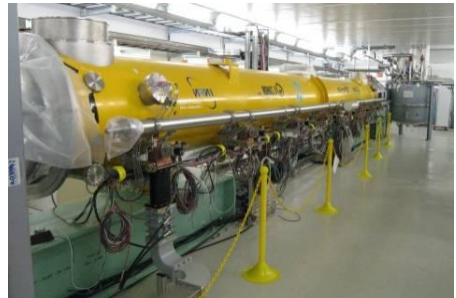

Fig. 10: The first cryomodule from DESY installed in the ASTA facility at Fermilab. This has eight 9-cell cavities.

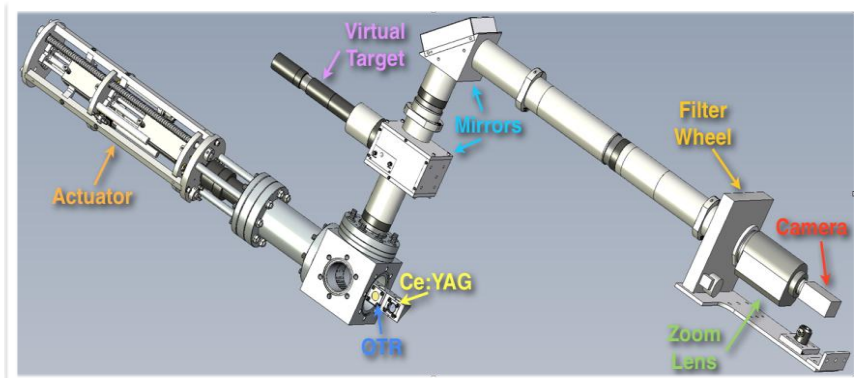

Fig. 11: A schematic of the prototype imaging station to be used at ASTA showing the three-position pneumatic actuator, screen holder, optical transport, filter wheel, and CCD camera.

## 4  Summary

We have described several of the issues and limitations one encounters with the imaging of relativistic electron beams. We have reported our initial tests at the A0PI facility and our plans to extend these studies to the GeV scale at the ASTA facility. We also have plans to test these concepts with 23-GeV beams at the FACET facility at SLAC in the coming year. It appears the future remains bright for imaging techniques in ILC-relevant parameter space.

## 5  Acknowledgments

The authors acknowledge the technical assistance of the A0PI project team on experiments and M. Ruelas and A. Murokh of RadiaBeam Technologies for technical discussions and prototype development.